# Progress in relativistic laser–plasma interaction with kilotesla-level applied magnetic fields

K. Weichman, 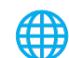 A. P. L. Robinson, 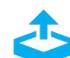 M. Murakami, 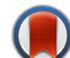 et al.

**COLLECTIONS**

Paper published as part of the special topic on Papers from the 63rd Annual Meeting of the APS Division of Plasma Physics

### ARTICLES YOU MAY BE INTERESTED IN

Microbubble implosions in finite hollow spheres
Physics of Plasmas **29**, 013105 (2022); https://doi.org/10.1063/5.0068815

Experimental investigations of hard x-ray source produced by picosecond laser-irradiated solid target
Physics of Plasmas **29**, 013107 (2022); https://doi.org/10.1063/5.0064541

Hot Raman amplification
Physics of Plasmas **28**, 062311 (2021); https://doi.org/10.1063/5.0049222

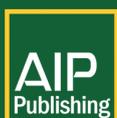

# Progress in relativistic laser–plasma interaction with kilotesla-level applied magnetic fields

K. Weichman,[1,a)] A. P. L. Robinson,[2] M. Murakami,[3] J. J. Santos,[4] S. Fujioka,[3] T. Toncian,[5] J. P. Palastro,[1] and A. V. Arefiev[6]

AFFILIATIONS

[1]Laboratory for Laser Energetics, University of Rochester, Rochester, New York 14623, USA
[2]Central Laser Facility, STFC Rutherford-Appleton Laboratory, Didcot OX11 0QX, United Kingdom
[3]Institute of Laser Engineering, Osaka University, Suita, Osaka 565-0871, Japan
[4]University of Bordeaux, CNRS, CEA, CELIA, UMR 5107, F-33405 Talence, France
[5]Institute for Radiation Physics, Helmholtz-Zentrum Dresden-Rossendorf e.V., 01328 Dresden, Germany
[6]Department of Mechanical and Aerospace Engineering and Center for Energy Research, University of California at San Diego, La Jolla, California 92093, USA

Note: This paper is part of the Special Collection: Papers from the 63rd Annual Meeting of the APS Division of Plasma Physics.
Note: Paper TI1 4, Bull. Am. Phys. Soc. 66 (2021).
a)Invited speaker. Author to whom correspondence should be addressed: kweic@lle.rochester.edu

ABSTRACT

We report on progress in the understanding of the effects of kilotesla-level applied magnetic fields on relativistic laser–plasma interactions. Ongoing advances in magnetic-field–generation techniques enable new and highly desirable phenomena, including magnetic-field–amplification platforms with reversible sign, focusing ion acceleration, and bulk-relativistic plasma heating. Building on recent advancements in laser–plasma interactions with applied magnetic fields, we introduce simple models for evaluating the effects of applied magnetic fields in magnetic-field amplification, sheath-based ion acceleration, and direct laser acceleration. These models indicate the feasibility of observing beneficial magnetic-field effects under experimentally relevant conditions and offer a starting point for future experimental design.

Published under an exclusive license by AIP Publishing. https://doi.org/10.1063/5.0089781

## I. INTRODUCTION

Recent advances in vacuum magnetic-field–generation techniques[1–6] have renewed interest in the fundamentals of laser–plasma interaction in the presence of strong magnetic fields. In part, this interest has been motivated by the potential for applied magnetic fields to benefit applications of laser–plasma interaction at relativistic intensity ($I_0 \gtrsim 10^{18}$ W/cm$^2$ for ~1-$\mu$m wavelength), including ion acceleration,[7–10] inertial fusion energy,[11–13] and the laboratory study of astrophysical phenomena.[14–17] From a basic science perspective, the understanding of the effects of strong magnetic fields in laser-generated plasma is still rapidly evolving, including the question of whether laser plasma is diamagnetic or generates and amplifies magnetic fields,[18–24] the effect of magnetic fields on electron transport in plasmas and the resulting ion dynamics,[7–10,25,26] and magnetic-field–associated changes in the direct laser acceleration of electrons.[27–30]

Plasma is conventionally considered diamagnetic and often acts to exclude magnetic fields; however, laser–plasma interactions have long been known to self-generate strong fields (e.g., inverse Faraday effect[31,32]) and amplify applied magnetic fields (e.g., flux compression[18]). The spatial localization of hot-electron production from an overdense target and the presence of a neutralizing cold return current offers additional opportunities for magnetic-field generation and amplification associated with kinetic electron dynamics,[20,23,24] among which is surface magnetic-field generation arising from the inability of the hot-electron current to change the applied field in a conductive opaque target.[21] This surface magnetic field can influence later plasma dynamics including target expansion[21] and may reverse the sign of the magnetic field generated by laser-driven implosions when it is destabilized.[33] The latter case is of particular interest as a platform for extreme magnetic-field amplification.[22] However, the process underlying the sign reversal phenomenon[33,34] is yet to be conclusively settled. In this work, we introduce a computationally efficient model for studying surface magnetic-field stability and demonstrate its ability to predict the sign of the magnetic field produced in implosions.

Until recently, the effect of applied magnetic fields on laser-driven plasma expansion and ion acceleration has primarily been studied in the context of astrophysical jet dynamics, involving long timescale (~nanosecond) evolution in sub-100-Tesla magnetic fields.[17,35,36] These studies have necessitated magnetohydrodynamic modeling, which eliminated the possibility of considering kinetic electron and ion dynamics. The sheath-based ion-acceleration regime driven by short, relativistic intensity laser pulses, on the other hand, is conducive to multidimensional kinetic modeling. Recent work in this regime has revealed the possibility of using an applied magnetic field to reverse the typical outward divergence associated with target normal sheath acceleration[37] into focusing and improving the ion energy and number.[7,10,38] In this case, ion focusing, which is highly desirable and much studied under non-magnetized conditions,[39–41] is produced by eventual magnetization of the electron sheath as the plasma expands.[10] However, observing ion focusing experimentally will require the spatial scale of the applied magnetic field to be comparable to or greater than the focal length. In this work, we introduce a simple scaling model for sheath magnetization and subsequent ion focusing, from which we predict that realistic ion focal lengths are likely compatible with the spatial extent of currently available applied magnetic fields.

While conventional electron acceleration mechanisms typically leave the majority of electrons cold either spectrally (e.g., wakefield-mediated acceleration[42,43]) or spatially (e.g., laser–solid, near-critical, and structured target interactions[44–46]) after the laser pulsed has passed, direct laser acceleration (DLA) with an applied magnetic field is capable of volumetrically heating electrons to relativistic energy.[27,28,30] In the regime where the applied magnetic field affects the acceleration dynamics in a single accelerating laser half-cycle,[29] even modestly relativistic laser pulses ($a_0 = 1$, where $a_0 = |e|E_0/m_e c \omega_0$ is the normalized peak laser amplitude for laser frequency $\omega_0$) can deliver significantly relativistic electron energy ($\gamma \gtrsim 10$ or more). A configuration employing a secondary laser pulse prior to the main accelerating pulse (to provide the preheating necessary to enter this regime) was recently demonstrated to heat the majority of electrons in a large plasma volume to nonperturbatively relativistic energy.[30] This is the first method to volumetrically generate relativistically thermal plasma at gas-jet–accessible density— conditions which are highly desirable for fundamental experimental studies in basic plasma physics,[47] astrophysics and laboratory astrophysics,[48–53] and laser-plasma physics.[54–57] In this work, we obtain an estimate for the average electron energy generated via magnetically assisted direct laser acceleration and evaluate its efficiency.

In this paper, we build on recent progress in three broad areas of relativistic laser–plasma interactions with kilotesla-level applied magnetic fields. In Sec. II, we present a simplified model for assessing the stability of the diamagnetic surface field produced by laser irradiation of an overdense plasma with an embedded magnetic field and demonstrate a correlation between surface magnetic-field instability and sign-reversed field amplification in laser-driven implosions. In Sec. III, we introduce a model for the length and time scales of sheath magnetization and ion focusing in sheath-based ion acceleration with an applied magnetic field and predict the magnetic-field length scale required for ion focusing to be observed experimentally. In Sec. IV, we formulate predictions for the electron energy and overall laser-to-electron energy conversion efficiency achievable from magnetically assisted direct laser acceleration in underdense plasma. In Sec. V, we summarize and discuss the implications of this work.

## II. STABILITY OF SURFACE MAGNETIC FIELDS AND FIELD AMPLIFICATION

In the conventional diamagnetic effect, electrons streaming through a magnetized region undergo momentum rotation, introducing currents that feed back on the magnetic field, reducing its amplitude. This feedback sets an upper limit on the electron current that can be obtained from momentum rotation, corresponding to the elimination of the applied magnetic field. In the case of laser–solid interaction, however, the conventional picture is altered by the presence of cold electrons within the solid, which act to prevent the magnetic field from changing. This screening effect allows the momentum rotation of laser-heated electrons to exceed the usual diamagnetic limit. These hot electrons then carry a net current past the solid–vacuum interface, driving a surface magnetic field with polarity opposite to the applied field. When the seed magnetic field ($B_0$) is perpendicular to the target-normal direction, the resulting surface field can exceed the applied field by as much as several orders of magnitude for a 1D-like laser pulse normally incident on an opaque target of thickness $\Delta x \lesssim r_L$, where $r_L$ is the Larmor radius.[21]

Although the process described above is a fundamental feature of laser–solid interaction with an embedded magnetic field, the conditions under which it is directly measurable are limited. The surface field exists only within the hot-electron sheath, which is comparable in size to the electron Debye length $\lambda_D = \sqrt{T_e/4\pi n_e e^2} \sim \lambda_0/2\pi$ (for laser wavelength $\lambda_0$, assuming the electron temperature $T_e \sim a_0 m_e c^2$ and density $n_e \sim a_0 n_c$, where $n_c = m_e \omega_0^2/4\pi e^2$ is the critical density).[21] Measurements of the magnetic field in the sheath region, while experimentally feasible,[58–61] remain challenging. In addition, the surface field must be comparable in magnitude to the azimuthal magnetic field produced by the finite laser spot size[59,62,63] in order to be visible, which requires a strong applied magnetic field, $B_0 \gtrsim a_0 m_e c^2/2|e|\Delta x$ (Ref. 21).

However, surface magnetic-field generation can also be indirectly observed through its effect on other processes. For example, target-transverse surface magnetic fields inhibit rear-surface ion acceleration[62] and for a very strong applied field (such that $\Delta x \gtrsim r_L$) favor ion acceleration from the front target surface.[21] In addition, the dynamics of the surface field can affect subsequent magnetic-field generation, such as the amplification of magnetic fields in imploding voids within an opaque target.[22,33,34,64] It was previously observed that the magnetic field at the inner surface of a laser-driven microtube target with an embedded magnetic field could become unstable, associated with density perturbation and eventual filamentation of the laser-driven surface, and, when appropriately timed, reach the center of the imploding void and change the sign of the magnetic field amplified by the implosion.[33]

In this section, we introduce a 2D planar configuration for studying surface magnetic-field destabilization through heating of the cold-electron population, which is enhanced by surface filamentation. The stability of the surface field in this planar configuration is found to predict sign reversal in microtube implosions conducted with analogous laser and target parameters [shown in Fig. 1, where planar cases Figs. 1(c) and 1(e) mock up implosion cases Figs. 1(d) and 1(f), respectively]. Our planar configuration consists of a few-micron-thick opaque target irradiated by one normally incident or two obliquely incident plane wave laser pulses with periodic transverse boundary conditions. Details of the simulation parameters are given in Subsection 1 of the Appendix.

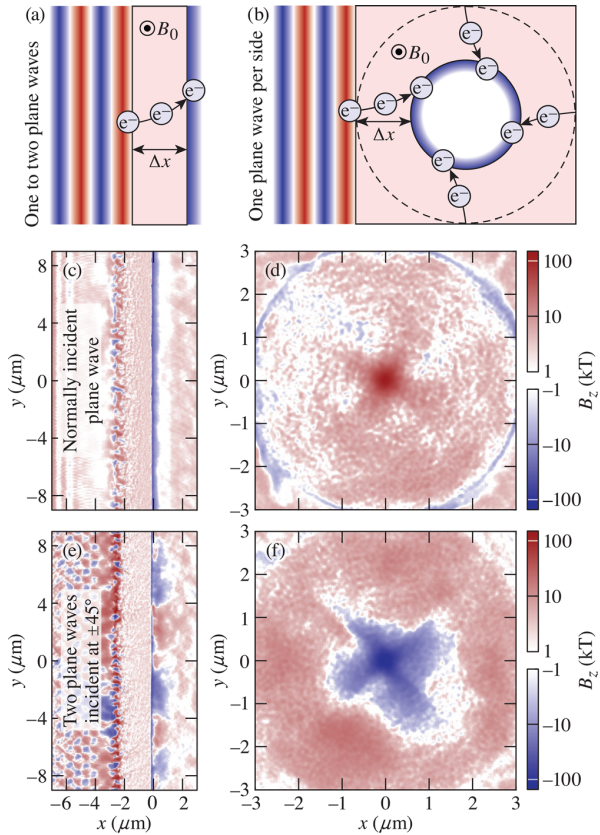

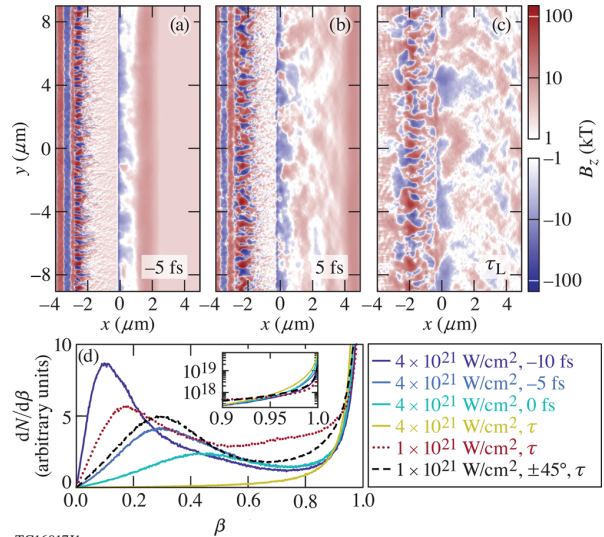

**FIG. 1.** Planar model capturing surface magnetic-field stability and sign of the amplified field in implosions. (a) and (b) Schematic of surface magnetic-field generation in (a) planar target and (b) implosion target with either square (solid line) or circular (dotted line) outer cross section. (c) Stable surface magnetic field in planar target with normally incident plane wave pulse ($I_0 = 10^{21}$ W/cm$^2$, $B_0 = 3$ kT) and (d) seed-aligned amplified magnetic field in square implosion target. (e) Unstable surface magnetic field in planar target with two obliquely incident pulses (each with $I_0$) and (f) amplified field in circular implosion target. Other parameters are given in Subsections 1 and 2 of the Appendix.

**FIG. 2.** Evolution of surface magnetic field and electron heating. (a)–(c) Time snapshots of magnetic-field evolution for a normally incident plane wave pulse with $4 \times 10^{21}$ W/cm$^2$. (d) Electron velocity distribution ($\beta = v/c$). Time is measured relative to when the peak of the pulse would reach the rear target surface. $\tau$ is the end of the pulse.

The irradiation of a thin opaque target by a normally incident plane wave laser pulse drives oscillation of the target surface and imposes periodic spatial modulation.[65–67] This surface modulation is unstable, leading to the growth of density and magnetic-field filaments into the target,[66–68] as can be seen for $-3\,\mu\mathrm{m} < x < 0$ in the time sequence shown in Figs. 2(a)–2(c). The growth of these filaments facilitates laser energy deposition within the target[66,69,70] and reduces the cold part of the electron population. In the limit of filament growth all the way through the target during the laser pulse duration, the cold population can be almost entirely eliminated [disappearance of the cold-electron peak in the $4 \times 10^{21}$ W/cm$^2$ sequence in Fig. 2(d)]. The cold-electron population can also be substantially reduced with only partial filamentation of the target when the laser is incident at an angle, especially in the presence of multiple interfering beams [e.g., $\pm 45^\circ$ case in Figs. 1(e) and 2(d), and analogous implosion in Fig. 1(f)], which significantly increases laser absorption.[71,72]

In a laser-irradiated opaque target with an embedded magnetic field, the cold-electron population carries transverse current that prevents diamagnetic reduction of the magnetic field by the hot electrons within the target bulk, and in doing so facilitates the generation of the surface magnetic field. The balance between hot- and cold-electron current breaks down when the majority of electrons in the target are heated, leading to the disruption of the surface magnetic field. An example of the disruption of the surface magnetic field as electrons are heated through target filamentation is shown in Figs. 2(a)–2(c). Although a distinct surface magnetic field is seen at early time, it breaks up as the filaments penetrate deep into the target. In conjunction with this breakup, the magnetic field several micrometers from the target surface changes sign from purely seed-field aligned to partly oppositely aligned [e.g., the blobs of negative field visible for $x > 0$ in Fig. 2(c)]. This change in the sign of the field tens of Debye lengths away from the surface, i.e., outside the sheath region, is only observed when the surface field is disrupted; otherwise it maintains the same sign as the applied field [cf., $x \gtrsim 0.5\,\mu\mathrm{m}$ in Figs. 1(c) and 1(e)].

In the context of a microtube implosion, whatever magnetic field is present when ions first reach the center of the target void will be amplified.[33] The stability of the surface magnetic field in the planar target configuration thereby functions as a proxy for the sign of the magnetic field that would be produced in a microtube implosion with similar parameters. To confirm the predictive capabilities of this model, we scanned over a number of parameters that affect electron heating and the growth of filaments in the target, including laser incidence angle (to represent targets with circular outer cross section and large spot size), laser intensity, pulse duration, seed magnetic-field strength, target thickness, and target density and composition. The parameters used for the planar model were chosen to match simulations of microtube implosions, covering variation in all the listed parameters.

In all cases, the predominant sign of the magnetic field a few micrometers from the rear target surface just after the end of the laser pulse in the planar model predicted whether a region of sign-reversed magnetic field was present in the imploded target. The planar model reproduces the same trends in magnetic-field sign as the imploding target studies in Refs. 22, 33, 34, and 64. For example, the case of a single laser pulse with intensity $I_0 = 1 \times 10^{21}$ W/cm$^2$ at normal incidence produces a stable surface magnetic field, whereas two pulses at 45° incidence drive the magnetic field unstable, which agrees with the result in Ref. 33 that implosions with a square outer cross section amplify the seed magnetic field, while those with circular outer cross section amplify the originally surface-generated field. Increasing the laser intensity beyond $2 \times 10^{21}$ W/cm$^2$ in the normally incident (square cross section-equivalent) case drives the surface magnetic field unstable (e.g., $4 \times 10^{21}$ W/cm$^2$ shown in Fig. 2); however, stability is recovered by increasing the applied magnetic field from 3 to 6 kT, in agreement with the seed-aligned field observed in the intensity scan in Ref. 64. Stability of the surface field in the planar model can also be recovered by increasing the target electron density (e.g., from 50 $n_c$ to 200 $n_c$), increasing the target thickness (e.g., from 3 to 6 $\mu$m), or decreasing the pulse duration (e.g., from 50 to 25 fs).

## III. SHEATH MAGNETIZATION

The ability of an applied magnetic field to inhibit electron motion across field lines, which restricts plasma expansion for a sufficiently strong target-transverse magnetic field,[21] is beneficial to ion acceleration when the magnetic field is aligned in the target-normal direction. In the limit of a very strong applied field ($\sim$10 kT), resonant electron acceleration ($\omega_{c0} \sim \omega_0$, where $\omega_{c0} = |e|B_0/m_e c$ is the cyclotron frequency)[27] has been shown to enhance radiation pressure acceleration.[8,9] However, 10-kT–level fields have a very high energy cost (the field energy scales as $B_0^2$) and are far from the currently realizable regime. At lower, experimentally relevant kilotesla- and 100 Tesla-level fields, applied magnetic fields increase the energy and number of ions accelerated by rear-surface sheath-based ion acceleration[7,10,38] and reverse the usual outward divergence characteristic of target normal sheath acceleration[37] into focusing about magnetic field lines.[10]

In this magnetized electron sheath acceleration, the applied magnetic field has beneficial effects in both the opaque target and the hot-electron sheath. In the target, the magnetic field assists in the transverse confinement of hot electrons, resulting in higher accelerating fields that drop more slowly in time compared to the unmagnetized case. This effect is most dramatic when the characteristic thickness of the target plus preplasma is comparable to or greater than the Larmor radius ($\Delta x \gtrsim r_L$) and is more readily observed in 3D simulations than in 2D due to the artificially slower drop-off of the accelerating sheath field in 2D (Ref. 10).

The applied magnetic field also magnetizes the sheath, enabling ion focusing. Magnetization of the sheath is not immediate, however, since the strong electric field associated with the hot-electron population initially dominates over the applied magnetic field. The initial period of ion acceleration in an unmagnetized sheath allows ions to develop an initial outward divergence (visible in the trajectories in Fig. 3), which later allows them to transversely overshoot the magnetized electron population, resulting in a focusing electric field. This focusing field eventually draws the ions back toward the laser axis, setting up oscillation of the highest-energy ions about the axis.[10]

In this section, we develop a simple estimate for the characteristic time and distance scales associated with magnetization of the hot-electron sheath and the ion focusing. An understanding of these scales is needed, for example, to determine the volume of magnetic field needed to observe ion focusing and evaluate the suitability of experimental platforms for magnetic-field generation. Magnetization of the sheath occurs when the magnetic field dominates the electric field in the equation of motion for hot electrons in the magnetic-field–transverse direction

$$\frac{d\mathbf{p}_\perp}{dt} = -|e|\mathbf{E}_\perp - |e|\frac{\mathbf{v}}{c} \times \mathbf{B}, \quad (1)$$

which requires at a minimum $E_\perp < B_0$.

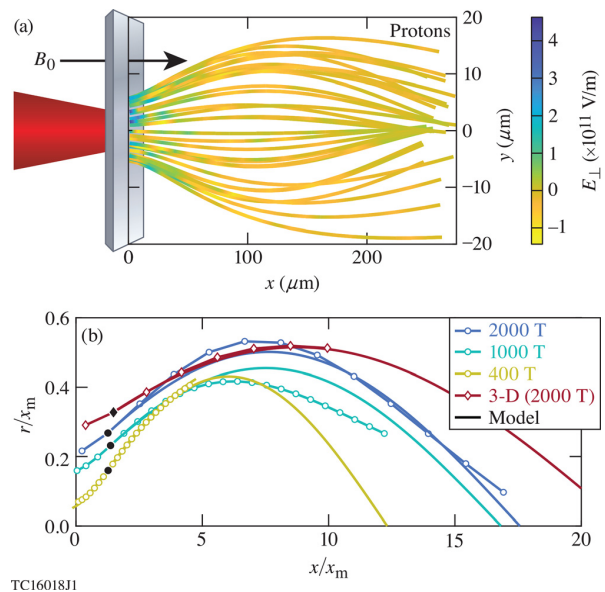

FIG. 3. Ion focusing in magnetized electron sheath acceleration. (a) Schematic of ion acceleration with a target-normal applied magnetic field with proton trajectories from the 1000-T 2D simulation (see Subsection 3 of the Appendix for parameters). (b) Proton trajectories averaged over all ions above the energies 10, 17, 25, and 4 MeV, respectively ($\sim$30% below the cutoff ion energy) from simulations scanning over magnetic-field strength while keeping the laser waist divided by $B_0$ constant. $x_m$ was calculated using these $\varepsilon_i$ and the initial $T_e$ [exp$(-\varepsilon_e/T_e)$ fit just after the peak of the laser pulse]. Thin lines are the solution to Eqs. (7) and (8) with the earliest magnetized points (black markers) as an initial condition and the best-fit constant obtained from the 3D case.

A simple estimate for the time and distance from the target surface where this transition occurs can be obtained from 1D sheath theory. Applying the isothermal Vlasov–Poisson–Boltzmann model for the expansion of a hot plasma slab, the magnitude of the longitudinal electric field drops in time as[73,74]

$$E(t) \approx \frac{E_0}{\omega_{\text{pi}} t}, \quad (2)$$

where $E_0$ is an initial value, $\omega_{\text{pi}} = \sqrt{4\pi e^2 n_0/M}$ is an effective ion plasma frequency for initial hot-electron density $n_0$ (in a multi-species plasma, $M$ corresponds to the lightest species[74]) and $\omega_{\text{pi}} t \gg \sqrt{2 \exp(1)}$ (Ref. 73). The transverse electric field has the same physical origin as the longitudinal electric field; therefore, it is reasonable to assume that the transverse electric field drops similarly to the longitudinal field in the initially unmagnetized sheath. The electric-field magnitude is linked to the thermal pressure of hot electrons, $E_0 \sim \sqrt{4\pi n_0 T_e}$, which gives

$$E(t) \sim \frac{\sqrt{MT_e}}{|e|t}. \quad (3)$$

The transition to magnetized behavior occurs when $E$ becomes comparable to $B_0$. From Eq. (3), the timescale for this to occur is roughly

$$t \sim t_{\text{m}} \equiv \frac{\sqrt{MT_e}}{|e|B_0}, \quad (4)$$

which is of the order of hundreds of femtoseconds for MeV electron temperature and kilotesla-level applied magnetic fields. Equation (4) suggests that subpicosecond pulses with moderate intensity have well-separated stages of ion energization and focusing, in agreement with simulations.

While applied magnetic-field sources[1,3–6,38,75] are typically long-lived (∼nanoseconds) relative to $t_{\text{m}}$, the spatial volume that can be magnetized may constrain their suitability for observing the focusing process. We convert $t_{\text{m}}$ to a simple distance scaling by assuming the velocity of the ion front $v_i = \sqrt{2\varepsilon_i/M}$ is approximately constant over $t_{\text{m}}$, where $\varepsilon_i$ is the ion energy cutoff. The distance from the target where the magnetization occurs is

$$x \sim x_{\text{m}} \equiv \frac{\sqrt{2\varepsilon_i T_e}}{|e|B_0}, \quad (5)$$

which is of the order of tens of micrometers for $\varepsilon_i \sim 10$ MeV, $T_e \sim 1$ MeV, and $B_0 \sim 1$ kT.

Following the magnetization transition, the inward-directed transverse electric field induces ion focusing. The length scale associated with focusing can be seen from the equations of motion for the protons at the ion front, which we consider using a test particle model. We assume the that the ion energy is fixed and nonrelativistic, such that $p_\perp \sim M v_i \sin\theta$, where $\theta$ is the characteristic divergence angle of the ions. The ion equations of motion can be written

$$\frac{d\sin\theta}{dv_i t} = \frac{|e|E_\perp}{2\varepsilon_i}, \quad (6)$$

$$\frac{dr}{dv_i t} = \sin\theta. \quad (7)$$

In the magnetized regime, the transverse electric field is associated with charge separation. We assume that the electric field is driven by the electron density, with negligible contribution from the small population of ions near the front. Approximating the electron density as uniform gives $E_\perp = -|e|n_e r/2$. As shown in Fig. 3(a), the $E_\perp$ experienced by ions remains approximately constant in $x$ after the magnetized transition until ions approach the axis, which is consistent with the near-constant $n_e$ observed in simulations. Assuming the initial electron density is given by $E \sim \sqrt{4\pi n_e T_e} \approx B_0$ at the onset of magnetization, the momentum equation can be written

$$\frac{d\sin\theta}{dt/t_{\text{m}}} \sim -\frac{r}{8\pi x_{\text{m}}}. \quad (8)$$

Equations (7), (8), and the initial condition $r \approx x_{\text{m}} \sin\theta_{\text{m}}$, where $\theta_{\text{m}}$ is the initial divergence angle, imply the natural time and length scales of the focusing process are of the order of $t_{\text{m}}$ and $x_{\text{m}}$, respectively. The solution to these equations is oscillatory ion motion about the axis, in agreement with the behavior observed in long duration simulations.

To probe the validity of $x_{\text{m}}$ as a scaling parameter, we conducted a series of particle-in-cell (PIC) simulations of magnetized electron sheath acceleration from a plastic (CH) target, varying the applied magnetic-field strength and the laser spot size, and the simulation dimensionality. Varying the laser spot size changes $T_e$ and $\varepsilon_i$, while changing the simulation dimensionality from 2D to 3D affects $\varepsilon_i$ alone. The details of these simulations are given in Subsection 3 of the Appendix.

Figure 3(b) shows averaged trajectories of high energy protons from the PIC simulations, where each point represents the average position of all protons within ∼30% of the cutoff ion energy. In all cases, the magnetized transition occurs at approximately $1.4 x_{\text{m}}$ [black markers in Fig. 3(b)]. We then use this point as an initial condition to solve Eqs. (7) and (8) [thin lines in Fig. 3(b)]. The observed ion trajectories are in reasonable agreement with the constant-density model for an initial perpendicular electric field at the ion location of $0.1 \sin\theta_{\text{m}} \sqrt{2\varepsilon_i/T_e} B_0$, where the factor $0.1 \sin\theta_{\text{m}}$ was obtained from the fit to the 3D trajectory. The agreement with the constant-density model is expected to be best for the 2000 T cases, which have the smallest value of $t_{\text{m}}$. Effects neglected in the model like time evolution of the electron density and temperature will cause the ion trajectory to be asymmetric about the maximum radius, as can be observed in the 1000 T case, and may result in somewhat longer ion focal length than predicted by the model.

The constant-density model nevertheless predicts ions return to the axis around 10-20 $x_{\text{m}}$, corresponding to hundreds of micrometers to millimeters for the electron temperatures, ion energies, and magnetic-field strengths that can be obtained experimentally. These distances are on a similar scale to the fields which can be created by state-of-the-art magnetic field generation techniques.[5,38] We therefore predict that observing ion focusing with an applied magnetic field is feasible, however spatial variation of the magnetic-field strength may need to be considered in order to accurately predict the ion focal length.

## IV. PLASMA HEATING BY MAGNETICALLY ASSISTED DIRECT LASER ACCELERATION

Electron acceleration by a relativistic plane wave laser pulse is conventionally reversible, leading to temporary energization of electrons in the pulse but no residual heating of the plasma. The reversibility of direct laser acceleration (DLA) can be broken in several ways.[27,43,76,77] For example, the reflection of a laser pulse from a sharp transition to overdense plasma non-adiabatically decouples electrons from the pulse at the critical density surface, allowing them to retain energy.[76] In underdense plasma, plasma-generated electric and magnetic fields break the usual invariants of electron motion.[43] However, both of these scenarios typically leave the majority of electrons sub-relativistic in either momentum or configuration space. The use of applied magnetic fields, on the other hand, can enable volumetric electron heating,[27,28] potentially to relativistic energies.[30]

Substantial effects of applied magnetic fields on direct laser acceleration may be observed in several regimes. First, electron acceleration is resonantly enhanced by a longitudinal magnetic field when $\omega_{c0} \gtrsim \omega_0$ (Ref. [27]), which corresponds to magnetic fields of the order of 10 kT for 1-$\mu$m laser wavelength. For weaker magnetic fields, electrons undergo momentum rotation on longer time scales than the laser period, under which conditions net acceleration can involve either the full pulse duration or a single laser half-cycle. In the former case, momentum rotation by a magnetic field transverse to the laser propagation direction changes the dephasing rate over the entirety of the pulse duration, resulting in the significant acceleration when the pulse duration is comparable to the rotation time.[28] In the latter case, an applied field aligned with the laser magnetic field alters the dynamics of electron acceleration during a single accelerating half-cycle.[29,30] The electron energy produced by magnetically assisted DLA in the half-cycle regime far exceeds that of the many-cycle regime for comparable laser energy; however, entering the half-cycle regime requires electrons to be preheated prior to the interaction.[30]

In Ref. [30], a short (tens of femtoseconds) laser pulse and a long (picosecond) laser pulse were combined to volumetrically heat a gas-jet–density plasma to multi-MeV average energy by successive stages of multi-cycle and half-cycle magnetically assisted DLA with a 100 Tesla-level applied field. The attractiveness of this result lies in the bulk-relativistic nature of the produced plasma, i.e., the property that more than half of the plasma electrons were heated to non-perturbatively relativistic energy ($\gamma > 2$) over a large volume. In this configuration, the preheating necessary to catalyze the half-cycle acceleration was provided by the short pulse.[30] From the standpoint of future experimental design, it is useful to consider whether the properties of heating by the long or the short pulse constrain the final electron energy.

In this section, we consider the dynamics of electron acceleration by half-cycle magnetically assisted direct laser acceleration in a preheated plasma. We demonstrate that the final energy of an electron is effectively independent of its starting energy above the preheating threshold, enabling a semi-empirical prediction for the average electron energy in this regime. Applying this single-electron model to a plasma, heating by the laser is found to be most efficient in a long, (relatively) high-density plasma with low laser intensity and high applied magnetic-field strength.

Direct laser acceleration in a plane wave imparts energy to electrons through work done in the laser polarization direction ($W = -|e| \int \vec{v} \cdot \vec{E} \, dt$). The laser magnetic field affects the energy gain process indirectly through rotation of the electron momentum. An applied magnetic field similarly rotates the electron momentum, but only affects the energy gain process if it can significantly change the direction of the momentum relative to the non-magnetized case. To illustrate the conditions under which the applied magnetic field performs significant momentum rotation during a single accelerating laser half-cycle, we write the electron equations of motion in terms of the angle the momentum makes with the forward (laser propagation) direction, $p_x = |p| \cos \theta$. With a $y$-polarized laser propagating in $x$ and a $z$-directed applied magnetic field,

$$\frac{d\theta}{ds} = \frac{|p_y|}{p_\perp} \frac{\frac{\omega_{c0}}{\omega_0} + \frac{1}{\beta} \left[ \cos\theta - \beta/\beta_\phi \right] \frac{da}{ds}}{\gamma \left[ 1 - (\beta/\beta_\phi) \cos\theta \right]}, \quad (9)$$

where $s = \omega_0 (t - x/v_\phi)$ is the laser phase variable with phase velocity $v_\phi$, $a$ is the normalized vector potential, $\omega_{c0} = |e|B_0/mc$ is the (non-relativistic) cyclotron frequency, $p_\perp = \sqrt{p_y^2 + p_{z0}^2}$ ($p_z$ is constant), $\beta_\phi = v_\phi/c$, and $\beta = |p|/\gamma mc$.

In Eq. (9), $d\theta/ds \approx 0$ minimizes the rate of change of the angle with respect to the electron slip in laser phase and thereby corresponds to the condition about which the majority of the electron acceleration occurs.[29,30] With an applied magnetic field, the angle given by $d\theta/ds = 0$ is only significantly different from the non-magnetized case when the electron already has energy prior to interaction with the laser pulse. It can be shown that the condition on the initial electron energy $\gamma_i$ required for the magnetic field to affect acceleration during a single half-cycle is[30]

$$\gamma_i \gtrsim f \sqrt{\frac{a_0}{2} \frac{\omega_0}{\omega_{c0}}} \equiv \gamma_0, \quad (10)$$

where $f$ is given by $f = \exp(-2a_0 f)$. $\gamma_0$ varies weakly with $a_0$. For $a_0 = 1$, $\gamma_0 = 0.3 \sqrt{\omega_0/\omega_{c0}}$.

When the initial electron energy exceeds $\gamma_0$, magnetically assisted DLA delivers a strong energy kick at the near-constant angle $\theta_m$ given by $d\theta/ds = 0$, with cyclotron-like electron orbits away from this condition [e.g., the calculated electron trajectory in Fig. 4(a)]. Eventually, cyclotron rotation returns the electron to $\theta_m$, facilitating another (positive or negative) energy kick.

The energy kick an electron receives depends sensitively on the starting energy and phase of the electron, with a maximum value of (Ref. [29])

$$\Delta \gamma \sim 2^{3/2} a_0^{3/2} \sqrt{\frac{\omega_0}{\omega_{c0}}}. \quad (11)$$

For a single kick, the energies before and after the kick are correlated [Fig. 4(b), from the solution to the equations of motion for a single electron in a plane wave]. However, this correlation disappears during subsequent kicks due to the sensitivity of the acceleration to the starting phase. When the laser pulse is long enough to deliver multiple kicks, the electron energy [e.g., black line in Fig. 4(b)] and its distribution [collected from varying the initial phase, black markers in Fig. 4(c)] become effectively independent of the starting energy $\gamma_i$. The condition on the laser pulse duration ($\tau$) for the electron to undergo more than one kick is $\tau \gtrsim \tau_L$, where $\tau_L$ is the maximum cyclotron period associated with a single kick $\tau_L \equiv \Delta \gamma \tau_{c0}$, where $\tau_{c0} = 2\pi/\omega_{c0}$ is the nonrelativistic cyclotron period, assuming the initial electron energy satisfies $\gamma_0 \lesssim \gamma_i \lesssim \Delta \gamma$.

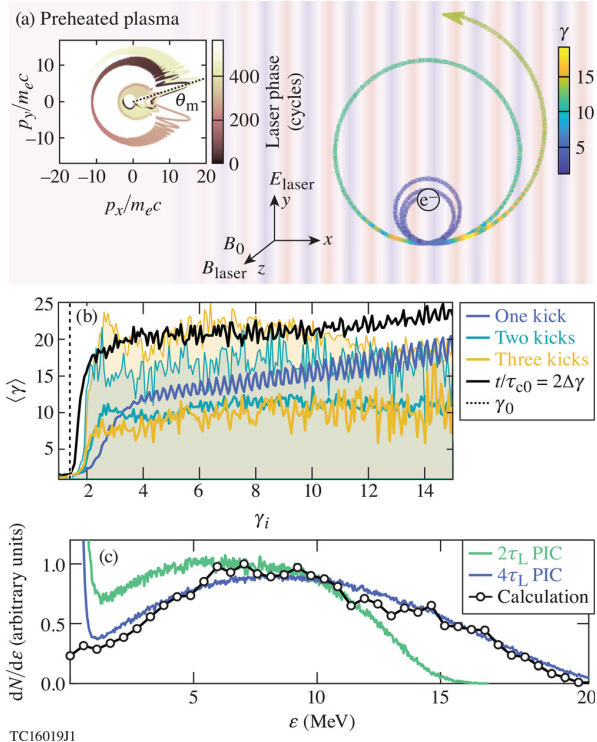

FIG. 4. Half-cycle magnetically assisted direct laser acceleration in a preheated plasma. (a) Example of many-kick electron acceleration process. (b) Phase-averaged electron energy from plane wave calculation vs initial electron energy $\gamma_i$. Shaded regions: 90% quantiles for the second and third kicks. (c) Electron energy spectrum from simulations with varying laser pulse duration and plane wave calculation (sampling $2 < \gamma_i < 5$ and initial phase). $a_0 = 1$ and $B_0 = 500\,\text{T}$; other parameters are given in Subsection 4 of the Appendix.

In the multi-kick regime, the characteristic energy of electrons following the interaction depends on the pulse duration. When the pulse duration is of the order of $\tau_L$ (e.g., picoseconds for 100 Tesla-level magnetic fields and $a_0 \sim 1$), a portion of electrons obtain sufficiently high energy during each kick that their cyclotron period exceeds the pulse duration, i.e., $\gamma \tau_{c0} > \tau$. These electrons, which represent a substantial fraction of the total population [including, e.g., the electrons lying above the shaded 90% quantiles in Fig. 4(b)], are thereby prevented from undergoing further energy gain or loss. The remaining lower-energy electrons have a shorter cyclotron period, allowing them to experience subsequent kicks, after which an additional portion will reach and retain high energy, further depleting the cold part of the population. Heuristically, this process results in a characteristic electron energy of the order of $\gamma \sim \tau/\tau_{c0}$, which is in good agreement with the electron energy obtained from the plane wave calculation.

Using the aforementioned properties of the acceleration process, it is possible to predict the average electron energy obtained from a laser pulse interacting with an underdense preheated plasma. For comparison with the vacuum theory, we conducted 1D PIC simulations of a Gaussian laser pulse interacting with a fully ionized hydrogen plasma with density $10^{-3}n_c$ with an initial waterbag energy distribution for the electrons (constant $dN/d\gamma$ below the cutoff). In addition to varying the laser pulse duration, intensity, and applied magnetic-field strength, the cutoff for the waterbag distribution was varied to control the fraction of electrons ($f_\text{hot}$) initially above $\gamma_0$. The electron energy distribution from PIC simulations is in good agreement with the phase-averaged vacuum plane wave calculation [Fig. 4(c)]. Additional simulation parameters are given in Subsection 4 of the Appendix.

The average electron energy retained in the plasma from each PIC simulation is shown in Fig. 5. Assuming the electrons that are left cold contribute negligibly to the average energy, the average energy is fairly well predicted by

$$\langle \gamma \rangle \approx 0.6 f_\text{hot} \frac{\tau}{\tau_{c0}}, \quad (12)$$

where the value 0.6 was obtained from a fit to the simulation data. The accuracy of the fit is somewhat degraded at high values of $f_\text{hot}$ due to the energization of protons in the plasma at the end of the simulation. This effect is especially pronounced for $4\tau_L$ case.

From Eq. (12), the overall efficiency of the energy gain process relative to the driving laser energy is

$$\eta = \frac{\varepsilon_\text{plasma}}{\varepsilon_\text{laser}} \approx 1.6 \frac{f_\text{hot}}{a_0^2} \frac{n_e}{n_c} \frac{L}{c\tau_{c0}}, \quad (13)$$

where $L$ is the length of the magnetized plasma. Assuming $f_\text{hot}$ is not very sensitive to the weak variation of $\gamma_0$ with $a_0$, the highest heating efficiency is therefore expected for small (relativistic) $a_0$, large $B_0$, and long plasmas.

Experimental magnetic-field generation platforms are currently capable of producing fields hundreds of Tesla strong over 100-$\mu$m to millimeter-scale distances, which corresponds to $L/(c\tau_{c0}) \sim 10$ to 100. Provided the majority of electrons are sufficiently preheated, Eq. (13) indicates the conversion efficiency can potentially reach percent-level for an electron density of the order of $10^{-2}$ to $10^{-3}n_c$ and $a_0 = 1$. Such a source of bulk-relativistic, optically diagnosable plasma is highly desirable for laboratory astrophysics, laser-plasma physics, and fundamental studies of relativistic effects in plasmas.

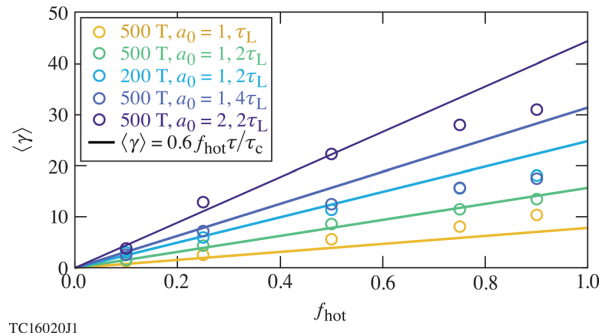

FIG. 5. Average electron energy from PIC simulations of magnetically assisted direct laser acceleration with varying initial fraction of electrons above $\gamma_0$ ($f_\text{hot}$). The plasma size is $2c\tau$, where $\tau$ is the FWHM pulse duration.

## V. SUMMARY AND DISCUSSION

In summary, we have presented simple models for evaluating the feasibility of observing the effects of applied magnetic fields in three different laser-plasma contexts. In laser–solid interaction, target-transverse applied magnetic fields trigger surface magnetic-field generation, influencing later plasma dynamics.[21] In laser-driven implosions,[22] the stability of the surface magnetic field was observed to correlate with the final sign of the amplified field.[33] The planar target model introduced in Sec. II facilitates evaluating the stability of the surface magnetic field and the sign reversal phenomenon of the amplified field in implosions, with significantly reduced computation cost. This model enabled observing the changes in electron heating that drive the surface magnetic field unstable, and predicted the sign of the magnetic field amplified in implosions. The planar model may also facilitate the design of future experiments to observe the sign reversal phenomenon.

In sheath-based ion acceleration, target-normal applied magnetic fields improve the accelerated ion energy and number, and produce ion focusing about the magnetic field lines.[7,10,38] The focusing phenomenon, in particular, is highly desirable to combat the typical outward divergence that otherwise reduces ion fluence far from the target surface. However, the applied magnetic field must remain sufficiently strong over the ion focal length in order to observe this effect. Section III presented simple scaling estimates for the ion focal length, and the time and distance scales for magnetization of the electron sheath, which initiates the change in ion divergence. These estimates indicate that current experimental magnetic-field generation capabilities should be suitable for observing the ion focusing effect, which occurs over a distance comparable to the produced magnetic field.

Finally, applied magnetic fields facilitate volumetric heating in underdense plasma via direct laser acceleration.[27,28] The strongest acceleration is observed when the magnetic field affects acceleration in a single half-cycle.[29,30] Although accessing this regime requires electrons to be preheated prior to the laser pulse,[30] its ability to produce plasma in which the majority of electrons are relativistic regardless of reference frame makes it highly attractive as a plasma heating mechanism. Section IV introduced an estimate for the average electron energy that can be produced from an initial preheated plasma, which increases with increasing pulse duration. Electron heating was found to be most efficient at low (but still relativistic) laser intensity and could potentially reach a few percent of the incident laser energy.

Together, these results highlight the promise of applied magnetic fields in relativistic laser–plasma interactions. Current magnetic-field capabilities can already enable novel and highly desirable phenomena relevant to laser-plasma applications. The continual development of magnetic-field–generation techniques supports these efforts by opening new parameter regimes to exploration.

## ACKNOWLEDGMENTS

This material is based upon work supported by the Department of Energy National Nuclear Security Administration under Award No. DE-NA0003856, the University of Rochester, and the New York State Energy Research and Development Authority. A.V.A. was supported by NSF Grant No. 1903098. The support of DOE does not constitute an endorsement by DOE of the views expressed in this paper. Particle-in-cell simulations were performed using EPOCH,[78] developed under UK EPSRC Grant Nos. EP/G054940, EP/G055165, and EP/G056803. Data collaboration was supported by the SeedMe2 project[79] (http://dibbs.seedme.org). This work used HPC resources of the National Energy Research Scientific Computing Center (NERSC), a U.S. Department of Energy Office of Science User Facility operated under Contract No. DE-AC02-05CH11231 using NERSC Award No. FES-ERCAP-0021627, and the Extreme Science and Engineering Discovery Environment (XSEDE),[80] which is supported by National Science Foundation Grant No. ACI-1548562, under Allocation No. TG-PHY210072 on the Texas Advanced Computing Center (TACC) at The University of Texas at Austin.

## AUTHOR DECLARATIONS
### Conflict of Interest

The authors have no conflicts to disclose.

## DATA AVAILABILITY

The data that support the findings of this study are available from the corresponding author upon reasonable request.

## APPENDIX: SIMULATION PARAMETERS

Simulations were conducted with the open-source particle-in-cell code EPOCH.[78] All simulations employed high-order cubic B-spline particle shape, which mitigates numerical heating[78] and delivers robust energy conservation.

### 1. Planar surface magnetic-field simulations

Planar simulations of surface magnetic-field generation were conducted in 2D using periodic boundary conditions and one or two 0.8-$\mu$m plane wave laser pulses with 50-fs duration (full width of $\sin^2$ profile in $|E|$) and a nominal intensity of $10^{21}$ W/cm$^2$. In the two-pulse case, the phase fronts were tilted at $\pm 45°$. The simulation domain was 18 $\mu$m in the transverse direction, which was varied without any qualitative changes in the magnetic-field profile. The simulation domain was resolved by 50 cells per laser wavelength in each direction. The nominal applied magnetic-field strength was 3 kT, oriented in the same direction as the laser magnetic field (out of the simulation plane). The plasma was fully ionized CH with a nominal electron density of 50 $n_c$ and thickness 3 $\mu$m, with 200 macroparticles per cell for electrons and 100 for each ion species. The magnetic-field snapshots shown were averaged over 5 fs.

## 2. Implosion simulations

Simulations of laser-driven implosions were conducted with four driving plane wave laser pulses incident on a microtube target (inner radius 3 $\mu$m) with either a square or circular outer cross section. The plasma, magnetic field, and initial laser parameters were matched to the planar target case, although higher spatial resolution (100 cells per laser wavelength) was used to resolve the imploded plasma near $r = 0$.

## 3. Ion-acceleration simulations

Ion-acceleration simulations were conducted in 2D and 3D with a laser pulse with peak intensity $2 \times 10^{19}$ W/cm$^2$, 150-fs (FWHM intensity) pulse duration, and 1.06-$\mu$m wavelength. The spot size was 3 $\mu$m in the cases with a 2-kT applied field and scaled with $2\,\mathrm{kT}/B_0$ for the other cases. The magnetic field was applied in the target normal (laser propagation) direction. The target was 5-$\mu$m-thick fully ionized CH with a 1.5-$\mu$m pre-plasma scale length [$n_e \propto \exp(-x/L_{\mathrm{pre}})$] with peak electron density 70 $n_c$. The resolution was $30 \times 30$ and $30 \times 20 \times 20$ cells per laser wavelength in 2D and 3D, respectively. Electrons (ions) were represented by 50 (25) and 10 (5) macroparticles per cell in 2D and 3D, respectively, with 150 and 20 ion macroparticles per cell within 0.5 $\mu$m of the rear target surface. The laser was polarized in the simulation plane in 2D.

## 4. Direct laser acceleration simulations

Direct laser acceleration simulations were conducted in 1D with a nominally 500 T magnetic field applied in the same direction as the laser magnetic field. The laser pulse was temporally Gaussian with a nominal pulse duration (FWHM in $|E|$) of $\tau = 2\tau_L$ (1.87 ps for 500 T), where $\tau_L = \Delta\gamma\tau_{c0}$ [$\tau_{c0}$ is the nonrelativistic cyclotron period, and $\Delta\gamma$ is given by Eq. (11)], nominal peak amplitude $a_0 = 1$, and 1-$\mu$m wavelength. The domain was spatially resolved by 40 cells per laser wavelength. The plasma was fully ionized hydrogen with density $10^{-3}\,n_c$ and thickness $L = 2c\tau$ (nominally 178 $\mu$m), with 100 (50) macroparticles per cell for electrons (protons). The initial electron distribution was initialized as a waterbag with constant $dN/d\gamma$ up to a cutoff value $\gamma_{\max}$, where $\gamma_{\max}$ was chosen to achieve the desired fraction of electrons above the preheating threshold $\gamma_0$ [Eq. (10), nominally 1.4]. The waterbag distribution was chosen to minimize the uncertainty in $f_{\mathrm{hot}}$ associated with the approximate nature of the prediction for the preheating threshold given in Eq. (10).